\documentclass[useAMS,usenatbib]{mn2e}
\usepackage{graphicx}
\title[Babylonian Lunar Appulses and Occultations]{Constraining $\Delta$T from Babylonian Lunar Appulse and Occultation Observations}
\author[G. Gonzalez]{Guillermo Gonzalez$^{1}$\thanks{E-mail:
ggonzalez@bsu.edu}\\
$^{1}$Ball State University, Department of Physics and Astronomy, 2000 W. University Ave., Muncie, 
IN 47306}
\begin{document}

\date{Accepted ??. Received ??; in original form ??}

\pagerange{\pageref{firstpage}--\pageref{lastpage}} \pubyear{??}

\maketitle

\label{firstpage}

\begin{abstract}
We examine in detail 15 Babylonian observations of lunar appulses and occultations made between 80 and 419 BC for the purpose of setting useful limits on Earth's clock error, as quantified by $\Delta$T, the difference between Terrestrial  Time and Universal time. Our results are generally in agreement with reconstructions of $\Delta$T using untimed solar eclipse observations from the same period. We suggest a revised version of the simple quadratic fit to $\Delta$T in light of the new results.
\end{abstract}


\section{Introduction}

Terrestrial Time (TT) is a uniform time variable essential for computation of the dynamics of Solar System bodies from gravitational theory. Universal Time (UT) is a nonuniform time system based Earth's rotation. The quantity $\Delta$T $(=$ TT-UT$)$ is calculated from the observed UT of an astronomical event, such as an eclipse or lunar occultation, and the predicted TT based on theory. Due to irregularities in changes of Earth's rotation, $\Delta$T cannot be predicted or extrapolated into the past from knowledge of the modern values. Reconstruction of historical changes in $\Delta$T, then, requires comparison of surviving records of carefully observed astronomical events to theoretical gravitational simulations.

\citet{ste16} provide the most up-to-date compilation of $\Delta$T values derived from historical astronomical observations. They employed lunar and solar eclipse observations made between 720 BC and 1620 AD and telescopic lunar occultation observations thereafter. \citet{ste12} had examined ancient Chinese lunar occultation observations going back to 69 BC, but they concluded that they were too unreliable to derive useful constraints on $\Delta$T. However, there does exist another as yet unexamined significant collection of ancient lunar occultation observations -- from Babylon. The Babylonian records describe several types of astronomical phenomena starting about the eighth century BC and continuing for some eight centuries. Their eclipse observations have been carefully analyzed by F. Richard Stephenson and his collaborators, but the occultation records have received scant attention.

We describe the observations and our methods in Section 2. In Section 3 we discuss the results. We present our conclusions in Section 4.

\section{Description of observations and methods}

Our primary source for the occultation observations is \citet{meis07}, which is a compilation of all the lunar occultations examined by Hermann Hunger and Abraham J. Sachs (e.g., \citet{sh01}) in the ancient Babylonian tables. De Meis lists a total of 41 events, which span the time period -675 to -79. His purposes in examining them is to check on the interpretations of Hunger and Sachs and to confirm the dates with modern astronomical calculations. In each case De Meis agrees with the dates assigned to the events by Hunger and Sachs, and he provides additional details and commentary.

In his examination of each event with modern astronomical calculations, De Meis adopts the $\Delta$T values from an unpublished manuscript by P.J. Huber. However, De Meis does quote the $\Delta$T value he employed in each case. In the present study, our approach is the reverse. Instead, we seek to constrain $\Delta$T for each well-dated event. For most of the events listed by De Meis, the dates are well-established from the information provided in the tablets, such as the year, month, and day relative to the start of the reign of a particular king. The older events, between -675 and -649, have more uncertain dates. In addition, for some events it is unlikely an actual observation is being described but rather a prediction.

In the following analysis, we exclude those events listed by De Meis that have uncertain dates. We also exclude those events that are likely to be mere predictions and those that are long duration, such as the passage of the Moon or a planet through a star cluster. An occultation of a bright star or planet, on the other hand, can be very useful in constraining $\Delta$T, given its short duration. However, even when an occultation report is available, it might not provide tight constraints on $\Delta$T. For example, some reports mention the culmination of a \textit{ziqpu} star \citep{sachs88} at the moment of the occultation. This permits us to know the UT of the event to within a narrow range.

As noted by \citet{ste12} grazing or near-grazing occultations are more useful in constraining $\Delta$T than non-grazing occultations when no other timing information is given. In addition, an occultation observed during twilight or when the Moon was close to the horizon can be used to set useful upper or lower bounds on $\Delta$T. Related to a grazing occultation is a close appulse. Some of these can also be useful in constraining $\Delta$T.

We examined each Babylonian record using the \textit{Skyfield} Python package.\footnote{http://rhodesmill.org/skyfield/}All the computations make use of the JPL DE422 ephemerides\footnote{ftp://ssd.jpl.nasa.gov/pub/eph/planets/README.txt}, which are valid between the years -3000 and 3000 \citep{jones11}. Topocentric apparent coordinates, including refraction, were computed for the relevant bodies over suitable ranges of TT and $\Delta$T for each event. In particular, for a given $\Delta$T, the TT of the appulse or occultation was determined. Several quantities were calculated for each combination of $\Delta$T and TT. The observers were assumed to be located in Babylon (long. = 42.42 deg E, lat. = 32.55 deg N). For occultations of bright stars, their proper motions were also included in the calculations.

In order to evaluate each occultation record properly, it is also necessary to determine the naked eye visibility limits. For this task, we make use of the detailed methods described in \citet{sch92}, supplemented with the equations in \citet{sch91} and \citet{sch93}. In brief, \citet{sch92} describe how to calculate the faintest magnitude visible to the naked eye when the Moon is the main glare source, such as during an appulse or occultation. Physically, the magnitude limits are set by scattered light from various background sources both in the atmosphere and in the eye. In addition to scattered moonlight, other contributing sources of background light include twilight, nighttime sky brightness, daylight (for a daytime occultation), and Earthshine. 

The only unknown adjustable parameter that can have a significant effect on the amount of scattered light, and thus on the limiting magnitude, is the extinction coefficient. \citet{sch93} notes that the visual extinction coefficient exhibits systematic seasonal variations, such that the summers tends to have higher extinction than the winters. Anecdotal evidence provided by the reviewer indicates that the weather is clearest in Baghdad (not far from the historical Babylon) from April to August. In this work we adopt visual extinction coefficient values of 0.2 for the winter, 0.3 for the summer, and 0.25 for the equinox months. However, we simply do not know what the actual extinction coefficient was at the time of each observation. This may not be as problematic for our analysis as it might seem, since some of the planets and stars involved are bright enough (and/or the Moon is faint enough) that extinction plays only a minor role in establishing the upper and lower bounds on $\Delta$T.

\subsection{Individual event descriptions}

In the following we describe the occultation records given by \citet{meis07} that set useful constraints on $\Delta$T. For each event, the proleptic Julian date given by the translators (and confirmed by De Meis) is given along with the translation and the original reference. Text within square brackets indicates the most likely translation of damaged regions of a tablet; when text is broken away it is indicated with ``[...]''.

\vspace{\baselineskip}
\noindent
\textbf{-79 September 23} -- Occultation of Antares\\
\noindent
``Night of the 3rd, beginning of the night,
alpha Scorpii entered the Moon.'' (Sachs and Hunger 1996, 485)

This is an early evening occultation of Antares by a four-day-old Moon. Only the immersion is mentioned. Antares has a visual magnitude of 0.91. The lower bound on $\Delta$T is about 8300 s. It is set by the visibility of Antares a few degrees above the horizon; the Sun is too far below the horizon for twilight to contribute significantly. The upper bound is about 11700 s for the assumed extinction. It is set by bright twilight. If the extinction coefficient is reduced to 0.2, then the upper bound increases to about 11900 s. We adopt $8300 < \Delta$T $< 11900$ s.\\

\noindent
\textbf{-84 January 10} -- Occultation of Jupiter\\
\noindent
``Night of the 16th, (when) Cancer [culminated,
Jupiter entered the Moon; ...; last part of the night, the Moon was
1/2 cubit...] rho Leonis, the Moon being 6 fingers high to the north,
it stood 2/3 behind Jupiter to the east.'' (Sachs and Hunger 1996, 465-467).

This is an occultation of Jupiter by a nearly full Moon. Cancer is said to culminate when the occultation occurs. Jupiter had a visual magnitude of -2.5; it should have been easy to see by the Moon. The cusp angle of the occultation was near 90 degrees, but it makes little difference for its visibility what precise value is adopted. Jupiter's altitude angle is near 60 degrees, and twilight does not contribute. There are four \textit{ziqpu} stars associated with Cancer: $\eta$ Cnc, $\theta$ Cnc, $\gamma$ Cnc, and $\delta$ Cnc. Cancer is not a very bright constellation. With the bright Moon, neither $\eta$ Cnc nor $\theta$ Cnc would have been visible. The culmination times for $\gamma$ Cnc and $\delta$ Cnc would have been very similar. $\delta$ Cnc would have culminated during Jupiter's occultation for $\Delta$T = 10000 s. 

It is difficult to decide how large an error to associate with this observation, since we do not know the precise methods used by the Babylonian astronomers. It is also not known how close to the actual culmination they would have considered to be close enough, but we have some clues. A change of 1000 s for $\Delta$T results in a change of 1.7 degrees for the altitude of $\delta$ Cnc. This should have been easily measurable and enough of a change in altitude to indicate the time of its culmination. Another clue is provided by the extensive Babylonian observations of lunar eclipses. \citet{ste97,ste06} provide detailed analyses of lunar eclipse timings and conclude that they were timed to the nearest
five degrees (equivalent to 1200 sec). For this occultation, this error in timing (determined by changing the RA of $\delta$ Cnc by 20 minutes), corresponds an error in $\Delta$T of about 800 s. There is reason to believe that the timing of occultations should be somewhat more accurate than the timing of lunar eclipses, but we will adopt this error as an upper bound. Thus, we adopt $9200 < \Delta$T $< 10800$ s for this event. Finally, we note that in this instance there is a discord between the transliteration and the translation by Sachs and Hunger as to where the square bracket ends.\\

\noindent
\textbf{-111 June 28} -- Occultation of Antares\\
\noindent
``[Night of the 10th, beginning of the night,
two]-thirds of the disk to the south, a third of the disk to the
north, alpha Scorpii [entered the Moon...]'' (Sachs and Hunger 1996, 345)

This is an another early evening occultation of Antares, this time by a ten-day-old Moon. This is a near grazing occultation. The Babylonian text mentions that one-third of the Moon's was disk to the north of the immersion point. This would have been a difficult observation; for the range of $\Delta$T values we explored, the minimum visibility magnitude was close to the magnitude of Antares. Assuming the scribe did see an actual occultation, and not merely an appulse, mistaken as an occultation, for instance, due to poor visibility, then the hard lower bound on $\Delta$T is 11500 s. The lower bound on $\Delta$T, assuming the limiting magnitude equals the magnitude of Antares, is 12500 s. Bright twilight sets a strong upper bound on $\Delta$T near 13500 s. For this range of $\Delta$T, Antares would have indeed disappeared near the Moon's northern limb. We adopt $11500 < \Delta$T $< 13500$ s, but we note that the constraint on the lower bound is weak.\\

\noindent
\textbf{-123 January 9} -- Occultation of Saturn\\
\noindent
 ``[Night of the 5th, when beta] Persei culminated,
[Saturn] entered the southern [horn] of the Moon.'' (Sachs and Hunger 1996, 277)

This is an early evening occultation of Saturn by a five-day-old Moon. It was a near grazing occultation. The scribe mentions that Saturn disappeared near the Moon's southern horn when $\beta$ Per culminated. The magnitude of Saturn was 0.52. The hard lower bound on $\Delta$T is about 12300 s, which is set by the requirement for an actual occultation. Twilight background was not a factor for this value of $\Delta$T. Saturn was nearly 2.5 magnitudes brighter than the limiting magnitude set by the Moon's glare, making this an easy observation. A hard upper bound on $\Delta$T of 14500 s is set by the need to be able to observe $\beta$ Per in the bright twilight. $\beta$ Per culminates at the moment of the occultation for $\Delta$T $= 13300$ s. The most probable bounds are $13000 < \Delta$T $< 14200$ s. We adopt $12300 < \Delta$T $< 14500$ s.\\

\noindent
\textbf{-124 September 4} -- Occultation of Venus\\
\noindent
``[Night of the 25th, last part of the night, the
Moon stood... behind Venus to] the east, (Venus) was set towards its 
northern horn, at 2/3 ? beru (= 20? deg) after sunrise, Venus entered
the northern horn of the Moon; it did not ... [...]'' (Sachs and Hunger 1996, 271)

The Moon is 26 days old. It is the only daylight occultation in De Meis's compilation. According to the scribe, it occurred 2/3 \textit{beru}, or one hour and 20 minutes, after sunrise; however, this number is not certain. The text mentions that Venus disappeared near the Moon's northern horn. The magnitude of Venus was -4.5. Over the range of $\Delta$T values considered, the limiting magnitude was near -3. Still, daylight observations of Venus are difficult. The scribe could have followed Venus from before sunrise until the occultation; its proximity to the Moon also would have aided its sighting. A value of $\Delta$T of 13700 s would correspond to the occultation occurring at the time given by the scribe. An error in the timing near $\pm$ 8 minutes would correspond to $\Delta$T spanning a range of 1000 s.\\

\noindent
\textbf{-132 October 28} -- Occultation of Regulus\\
\noindent
``Night of the 21st when Cancer culminated, 2?[...]
one-third of the disk to the north, alpha Leonis
entered the Moon; last part of the night, the Moon was 2 fingers
behind alpha Leonis, it was near.'' (Sachs and Hunger 1996, 215-217)

\noindent
``Night of the 21st, last part of the night, when Cancer culminated,
[one third of the disk]
[to the north alpha Leonis ] entered [the moon]; when the four stars
of the breast culminated, [it came out] from the moon[....].'' \citep{meis07}

This is an occultation of Regulus by a 22-day-old Moon near the end of the night. The scribe is very detailed in describing the circumstances. Both immersion and emersion are mentioned. In addition, it is noted that Cancer culminated at immersion and the stars of the breast of Leo culminated at emersion. From these descriptions we can obtain two independent sets of constraints on $\Delta$T.  

At immersion Regulus should have been easily visible beside the Moon's limb. Twilight background was not a factor. The four \textit{ziqpu} stars of Cancer cross the meridian over a time span of 13 minutes, but $\gamma$ Cnc and $\delta$ Cnc (which span a narrow range in RA) are much brighter than the other two.  At immersion, $\gamma$ Cnc culminates for $\Delta$T $= 11200$ s. Taking into account the likely uncertainty in timing the transit of Cancer, we adopt $10700 < \Delta$T $< 11700$ s for the immersion. 

The stars of the breast of Leo include Regulus, $\zeta$ Leo, $\gamma$ Leo, and $\eta$ Leo. At emersion, $\zeta$ Leo culminates for $\Delta$T $= 10800$ s. Morning twilight is an important factor for the visibility of $\zeta$ Leo; it would not have been visible for $\Delta$T less than 10800 s. $\gamma$ Leo would have been visible for $\Delta$T as small as 10300 s, but then all four stars of the breast would not have been visible. The four stars of the breast of Leo cross the meridian over a time span of 11 minutes; this permits the upper bound on $\Delta$T to be as large as 10500 s. Thus, we adopt $10800 < \Delta$T $< 11500$ s for the emersion. Increasing the extinction coefficient would have no effect on the immersion and would slightly increase the lower bound on the emersion $\Delta$T value. The internal consistency of the details of these events reassures us that they are reported reliably. For the full set of observations, we adopt $10800 < \Delta$T $< 11500$ s.\\

\noindent
\textbf{-226 June 5} -- Occultation of Jupiter\\
\noindent
``[...Night of the 5th beginning of the night,
the Moon was] 1 1/2 cubits behind theta Leonis, 4 fingers in front of
Jupiter, it was set towards its northern horn.'' (Sachs and Hunger 1989, 135)

De Meis assumes this is a report of an observed occultation. However, the description follows more closely the textual formula for a predicted occultation; it begins by giving the angle between the two objects in question and then closes with ``it was set towards.'' If this is a report of an early evening occultation of Jupiter by a five-day-old Moon, then we can apply the following analysis. 

We are told that Jupiter's immersion occurred near the Moon's northern horn; it was a near-grazing event. Jupiter had a visual magnitude of -1.9; it should have been be easy to see beside the Moon. A hard lower bound of 12700 s for $\Delta$T is obtained by requiring that the event was an actual occultation instead of an appulse. Twilight background was not a factor. As a result of the parallax effect of the observer's changing perspective on the Moon relative to the background stars, increasing $\Delta$T causes the immersion spot on the Moons's limb to move south from the northern horn. Keeping the immersion spot near the northern horn restricts the upper bound on $\Delta$T to about 13000 s. Thus, we would adopt $12700 < \Delta$T $< 13000$ s for this event under the assumption it was an observed occultation. But, as noted above, it is unlikely the occultation was actually observed.\\

\noindent
\textbf{-283 February 22} -- Occultation of Antares\\
\noindent
``Night of the 19th, last part of the night,
alpha Scorpii entered the Moon; clouds were in the sky.'' (Sachs and Hunger 1988, 301)

This is an occultation of Antares near the end pf the night by a 20-day-old Moon. This is a near grazing occultation. This would have been a difficult observation. The minimum visibility magnitude was just a few tenths of a magnitude fainter than Antares. Assuming the scribe did see an actual occultation, and not merely and appulse, which was mistaken as an occultation due to poor visibility (since the scribe mentions the presence of clouds), then the hard upper bound on $\Delta$T is 15000 s. No lower bound can be set for this event.\\

\noindent
\textbf{-284 October 18} -- Occultation of Mars\\
\noindent
``Night of the 11th, beginning of the night, the
Moon was [...] in front of Mars [...] [...] one-third of the disk
towards the north, Mars entered the Moon.'' (Sachs and Hunger 1988, 295)

This is a middle night occultation of Mars by a 12-day-old Moon. We are also told that one third of the Moon's disk was to the north of the immersion point. Mars was bright with a visual magnitude of -1.4; it should have been easy to see beside the Moon's limb. We can set a hard upper bound of 14500 s for $\Delta$T by the requirement for an occultation. The observation of the immersion point on the limb sets the lower bound at 11000 s. Thus, we adopt $11000 < \Delta$T $< 14500$ s for this event.\\

\noindent
\textbf{-289 June 11} -- Occultation of Saturn\\
\noindent
``Night of the 4th, be[ginning of the night...]
when [...] culminated, Saturn entered the [northern?] horn of the
Moon.'' (Sachs and Hunger 1988, 277)

This is an occultation of Saturn by a 5-day-old Moon. Saturn had a visual magnitude of 1.2. The scribe mentioned that the immersion point was near the northern horn. Although the scribe notes that a $\textit{ziqpu}$ star culminated when Saturn was occulted, its name is not readable on the cuneiform tablet. De Meis offers the identification as $\theta$ Oph. This seems reasonable, as there aren't any other $\textit{ziqpu}$ stars near culmination for a reasonable range in $\Delta$T.  Twilight background was not a factor in constraining $\Delta$T. The lower bound on $\Delta$T of 12900 s is set by the visibility of Saturn at low altitude. $\theta$ Oph culminates when Saturn is occulted for $\Delta$T $= 14200$ s. The upper bound is somewhat weakly constrained by the requirement that the immersion point is near the northern horn.  Thus, we adopt $12900 < \Delta$T $< 15000$ s for this event.\\

\noindent
\textbf{-289 June 10} -- Occultation of Regulus\\
\noindent
``[Night of the 3rd], beginning of the night,
alpha Leonis entered the Moon.'' (Sachs and Hunger 1988, 277)

This is an occultation of Regulus by a 4-day-old Moon. This event lacks details, but since it occurs at low altitude after sunset, we can set a useful bound based on visibility. We find that $\Delta$T must be greater than 12500 s for Regulus to be visible near the Moon's limb when it is only a few degrees above the horizon. Twilight background is not a factor. Thus, we adopt $\Delta$T $> 12500$ s.\\

\noindent
\textbf{-328 September 30} -- Occultation of Aldebaran\\
\noindent
``[Night of the 17th, last part of the night,]
[2/3 of the lunar disk to the no]rth, one-third of the lunar disk to
the south, alpha Tauri came out from the Moon [...].'' (Sachs and Hunger 1988, 185-187).

This is an emersion of Aldebaran from an 18-day-old Moon. The scribe notes that Aldebaran emerged at a point on the limb with one-third of the disk to its south. This would have been a challenging observation, even though twilight background would not have been a factor. The limiting magnitude at the Moon's limb is very close to the average magnitude of Aldebaran, 0.86. However, the visual magnitude of Aldebaran ranges from 0.75 to 0.95 magnitudes. The requirement of an occultation, as opposed to an appulse, gives us a hard lower bound for $\Delta$T of 13600 s. Even if the moment of emersion was not visible, the mention of the location of Aldebaran near the Moon's disk implies the scribe saw it at least shortly after emersion. Taking into account the location of the emersion point, we adopt $13600 < \Delta$T $< 17000$ s.\\

\noindent
\textbf{-416 July 21} -- Appulse of Saturn\\
\noindent
``Year 7. Month IV, the 22nd, 64 degrees before sunrise,
Saturn came out from the northern horn of the Moon.'' (Sachs and Hunger 2001).

Although the scribe describes this event as an emersion, it was actually an appulse. No reasonable values of $\Delta$T result in this being an occultation. Saturn had a visual magnitude of 0.1. Since the event occurred near the middle of the night, twilight background was not a factor. The ``emersion'' is described as occurring 64 degrees ($=$ 4$^{\rm h}$ 16$^{\rm m}$) before sunrise. The nominal value of $\Delta$T corresponding to the closest approach, given this timing, is about 15000 s. Although Saturn should have been bright enough to see by the limb of the Moon at closest approach, the scribe believes he has seen an emersion. Given this, we don't actually know whether Saturn was observed at its closest approach to the Moon or some time after. Assuming it was observed 15 minutes after the closest approach, then $\Delta$T $= 15800$ s. If we assume an additional error of $\pm15$ minutes in estimating the length of time elapsed between the event and sunrise, we arrive at $14000 < \Delta$T $< 16500$ s.\\

\noindent
\textbf{-416 September 1} -- Appulse of Mars\\
\noindent
``Year 7. Month VI, the 5th, first part of the night,
Mars was 1 finger below the southern horn of the Moon, it came close.'' (Sachs and Hunger 2001).

This event was indeed an appulse for all reasonable values of $\Delta$T. One finger corresponds to 5 arc minutes; the closest distance from the Moon's limb was actually about 19 arc minutes. Mars had a visual magnitude of 0.7. It occurred in the early evening with the Moon low in the west. A lower bound on $\Delta$T of 14500 s is set by the requirement for the visibility of Mars near the Moon when its altitude was very low. We assume a visual extinction coefficient of 0.25; twilight background was not a factor. Changing the extinction coefficient to 0.3 increases $\Delta$T by about 200 s. We adopt $\Delta$T $> 14300$  s.\\

\noindent
\textbf{-418 June 19} -- Appulse of Venus\\
\noindent
``Night of the 25th, last part of the night,
Venus came close to the southern horn of the Moon.'' (Sachs and Hunger 1988).

This is an close appulse of Venus with a nearly new Moon, observed before sunrise. Venus had a visual magnitude of -4.5. The fact that the scribe describes the event as a close appulse instead of an occultation is highly significant. Venus should have been bright enough to see near the Moon's limb over a wide range of altitudes and twilight sky brightnesses. The phrase ``came close to the southern horn'' implies that the scribe observed the moment of closest approach of Venus to the Moon's limb, which occurs near the horn (or cusp). Any disappearance of Venus due to an occultation would have been obvious. This requirement of an appulse, as opposed to an occultation, gives us a hard upper bound for $\Delta$T of 15750 s. A hard lower bound of 12900 s for $\Delta$T is set by the requirement that the event occur prior to sunrise, so that it is still considered to occur during the last part of the night. Changes in the extinction coefficient do not have any significant effects on these bounds, but anomalous refraction can alter the lower bound. Thus, we adopt $12900 < \Delta$T $< 15750$ s.\\

For the three appulses we can assume they are effectively simultaneous and arrive at the following bounds for the year -417: $14300 < \Delta$T $< 15750$ s. We summarize the results from this section in Table 1.

\begin{table*}
\centering
\begin{minipage}{160mm}
\caption{Allowed $\Delta$T ranges for each Babylonian occultation observation discussed in the text.}
\label{xmm}
\begin{tabular}{lccl}
\hline
Julian date & $\Delta$T range & Object occulted & lunar phase\\
 & (s) & &\\
\hline
-79 September 23 & 8300-11900 & Antares & 0.14+\\
-84 January 10 & 9200-10800 & Jupiter & 0.97-\\
-111 June 28 & 11500-13500 & Antares & 0.81+\\
-123 January 9 & 12300-14500 & Saturn & 0.25+\\
-124 September 4 & 13200-14200 & Venus & 0.14-\\
-132 October 28 & 10800-11500 & Regulus & 0.52-\\
-226 June 5 & 12700-13000 & Jupiter & 0.27+\\
-283 February 22 & $<$15000 & Antares & 0.68-\\
-284 October 18 & 11000-14500 & Mars & 0.90+\\
-289 June 11 & 12900-15000 & Saturn & 0.23+\\
-289 June 10 & $>$12500 & Regulus & 0.14+\\
-328 September 30 & 13600-17000 & Aldebaran & 0.91-\\
-416 July 21 & 14000-16500 & Saturn & 0.52-\\
-416 September 1 & $>$14300 & Mars & 0.30+\\
-418 June 19 & 12900-15750 & Venus & 0.09-\\
-417 combined & 14300-15750 & --- & ---\\
\hline
\end{tabular}
\end{minipage}
\end{table*}

\section{Discussion}

We have been able to obtain constraints on $\Delta$T, the correction to Earth's rotation clock, from 14 Babylonian lunar appulse and occultation reports (not counting the -226 occultation). The results from Table 1 are shown in Figure 1. In addition, upper and lower bounds on $\Delta$T derived by \citet{ste16} from four untimed total and annual solar eclipses are included in the figure. The untimed solar eclipses were crucial for \citet{ste16} to fix the temporal behaviour of $\Delta$T.

There are two important points to notice. First, the general increasing trend of $\Delta$T backwards in time is similar in the two datasets. Second, while the average range of allowed $\Delta$T is similar in the two datasets, one occultation has a smaller range than any eclipse from this time period. This is the occultation of Regulus in -132. The bounds it sets on $\Delta$T are consistent with the bounds set by the nearly simultaneous -135 total solar eclipse. Between the two, the formal allowed range for $\Delta$T is 11220 to 11300 s.

It needs to be emphasized that the ranges plotted in Figure 1 are not error bars, but are lower and upper bounds. Of course, uncertainties can be associated with each upper and lower bound, but in most cases they should be small ($\sim 100-200$ s). The bounds have varying degrees of certainty. For the events based on the transit timings of $\textit{ziqpu}$ stars, the bounds are somewhat softer than those based on near grazing occultations and appulses.

Also included in Figure 1 is a simple parabolic fit to $\Delta$T (equation 4 in \citet{ste16}), only taking into account the lunar appulse and occultation data. The appulse and occultation data are generally consistent with the curve. However, the fit to the events can be improved if the curve is shifted downward by about 500 s. In fact, \citet{ste16} calculated a spline function correction to the quadratic fit (see their Figure 15), which does provide a better overall fit to our constraints. Our data are also more consistent with a smaller quadratic coefficient, as shown in Figure 1.

The cluster of events near the year -120, however, appears to be discordant. Such a large and rapid change in $\Delta$T over just a few decades is highly implausible. If it were just one outlier event, it could be dismissed as a prediction, rather than an actual observation, or perhaps containing scribal errors. However, the three events appear to corroborate each other, and they involve occultations of three different objects. In the case of the year -111 event, its lower bound is closer to the mean trend line than the other two slightly earlier events; we conclude that this event is not really discordant.

Babylonian astronomers were also located in Nineveh and Uruk. To test the possibility that the events of the years -123 and -124 were obtained from one of these two locations, we repeated the analyses from the previous section. Uruk makes the disagreement even worse for the -123 Saturn occultation. However, placing the observer in Nineveh for this event changes the hard lower bound on $\Delta$T to 9700 s (in order to be an occultation). $\beta$ Per culminates for $\Delta$T = 12700 s at the moment of the occultation. However, for Saturn to be seen disappearing near the Moon's southern horn, a $\Delta$T value less than 11000 s is preferred. We don't have any evidence that this observation was made at Nineveh, however.

For the year -124 daylight occultation of Venus, placing the observer in Nineveh increases $\Delta$T by about 1000 s. Moving to Uruk only changes $\Delta$T by about 200 s. Thus, this event is still problematic, even allowing that Babylon might not have been the place of observation. \citet{ste11} notes that discordant $\Delta$T values derived from Babylonian lunar eclipse observations are not uncommon. In addition, the discordant values are all based on timings relative to sunrise or sunset; timings based, instead, on culminations of $\textit{ziqpu}$ stars are generally more self-consistent. The timing of the daylight Venus occultation is made relative to sunrise. What's more, this is the only daytime occultation in De Meis's list. Lastly, the place on the tablet where the timing of this event is given appears to be damaged. For these reasons we conclude that the report of this event is unlikely to be reliable.

In summary, there are good reason to treat the year -124 occultation as less reliable than the other events in Table 1. The lower bound of the year -111 occultation is consistent with the mean trend. Only the year -123 occultation is (mildly) discordant. The lower bound is based on the assumption that the event was an occultation rather than an appulse. Relaxing this assumption (e.g., due to poor transparency) can reduce the lower bound by a modest amount, but the mention of the immersion occurring near Moon's southern horn implies it was at least a close appulse.

\begin{figure}
\includegraphics[width=3.5in]{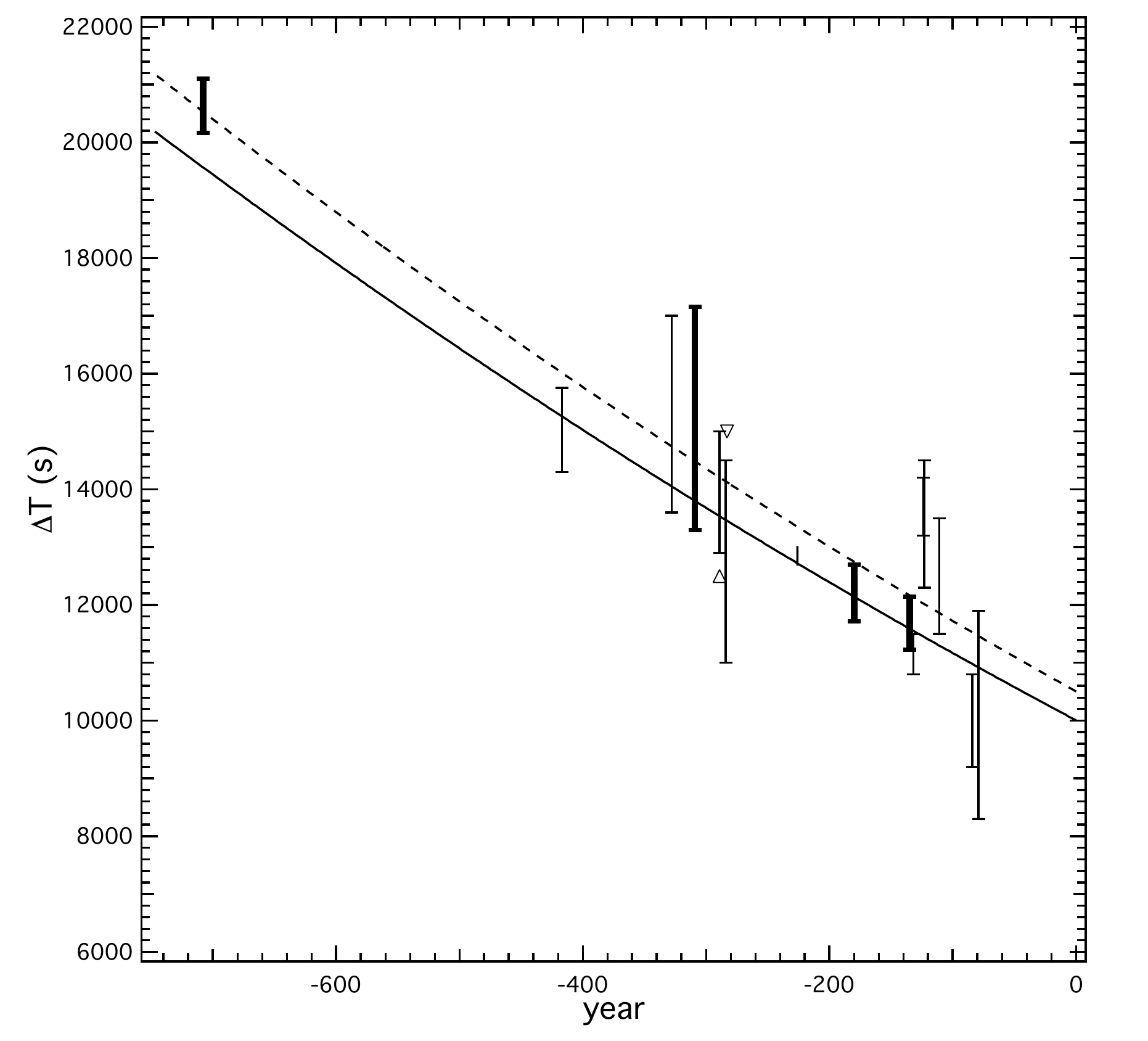}
\caption{Results for $\Delta$T from the Babylonian occultation observations (thin bars). The four thick bars represent $\Delta$T values from high quality untimed total and annular solar eclipses \citep{ste16}. Each bar represents the full range of allowed $\Delta$T for a given event. The downward pointing triangle is an upper bound, while the upward pointing triangle is a lower bound. The short vertical line corresponds to the event in -226. The dashed curve represents the simple quadratic fit from equation 4.1 of \citet{ste16}, which has a coefficient of 32.5 s~cy$^{\rm -2}$. The solid curve was calculated using a quadratic coefficient of 31.0 s~cy$^{\rm -2}$.}
\end{figure}

\section{Conclusions}

We have demonstrated that 14 Babylonian records of lunar appulses and occultations from the first to the fifth centuries BC provide important constraints on Earth's clock error, $\Delta$T. All the events we examined, except for two near 120 BC, are consistent with previous research done on $\Delta$T variations determined from solar eclipse observations from the same period. One event, in 133 BC, provides constraints on $\Delta$T comparable to the best ancient solar eclipse records.

Improved constraints on ancient variations in $\Delta$T can have substantial scientific payoffs. First, we can apply a revised trend in $\Delta$T to determine more precisely the observational circumstances of astronomical events as far back as the eighth century BC. Second, we can be more confident in extrapolating $\Delta$T to earlier times. Third, we can improve our understanding of geophysical processes that cause changes in Earth's rotation (e.g., \citet{mit15,hay16}).

\section*{Acknowledgments}

We thank the reviewers for very helpful comments and suggestions.

\bsp

\label{lastpage}

\end{document}